\documentclass{article}

\usepackage{spconf,amsmath,graphicx}
\usepackage{times}
\usepackage{soul}
\usepackage{url}
\usepackage[hidelinks]{hyperref}
\usepackage[utf8]{inputenc}
\usepackage[small]{caption}
\usepackage{graphicx}
\usepackage{amsmath}
\usepackage{amsthm}
\usepackage{booktabs}
\usepackage{algorithmic}
\usepackage[ruled,vlined]{algorithm2e}
\usepackage{diagbox}
\usepackage{cancel}
\usepackage{multirow}
\usepackage{comment}
\usepackage{mathtools}
\usepackage{amsbsy}
\usepackage[utf8]{inputenc}
\usepackage{dirtytalk}
\usepackage[dvipsnames]{xcolor}

\newcommand{\figref}[1]{Figure~\ref{#1}}

\newcommand{\tabref}[1]{Table~\ref{#1}}

\title{PROGRESSIVE IMAGE SUPER-RESOLUTION VIA NEURAL DIFFERENTIAL EQUATION}
\name{Seobin Park and Tae Hyun Kim}
\address{Hanyang University, Seoul, South Korea \\\{seobinpark, taehyunkim\}@hanyang.ac.kr}
\begin{document}
%
\maketitle
\begin{abstract}

\noindent We propose a new approach for the image super-resolution (SR) task that progressively restores a high-resolution (HR) image from an input low-resolution (LR) image on the basis of a neural ordinary differential equation. In particular, we newly formulate the SR problem as an initial value problem, where the initial value is the input LR image. Unlike conventional progressive SR methods that perform gradual updates using straightforward iterative mechanisms, our SR process is formulated in a concrete manner based on explicit modeling with a much clearer understanding. Our method can be easily implemented using conventional neural networks for image restoration. Moreover, the proposed method can super-resolve an image with arbitrary scale factors on continuous domain, and achieves superior SR performance over state-of-the-art SR methods.

\end{abstract}

\section{Introduction}

Image super-resolution (SR) is a classic low-level vision task that aims to recover a high-resolution (HR) image from a given low-resolution (LR) input image. For several decades, a large volume of literature documents the high demand of SR technique in various vision applications. However, SR problem still remains a challenge and is difficult to solve because it is a highly ill-posed inverse problem.

With the recent development of deep learning technology, numerous deep-learning-based SR methods~\cite{srcnn,vdsr,rdn} have been presented, and they have shown plausible results. 
To further improve the SR performance, many researchers have attempted to restore the high-quality image by recovering the fine details of the LR input image progressively~\cite{dbpn,srfbn}. 
Many previous works hinged on this progressive SR procedure are based on a variant of feedback network in the human visual system~\cite{zamir2017feedback}, and they show satisfactory SR results. 
However, owing to lack of theoretical clarity on the progressive system, these approaches need to develop a well-engineered method. 
For example, the number of iterations for the gradual refinements~\cite{drrn} and complicated learning strategies~\cite{srfbn} as well as the network architectures~\cite{dbpn,prlsr} are considered to improve the SR performance.
Several researchers have conducted studies on differential equations to solve the image restoration problems~\cite{tnrd,blind_pde}. They also have developed progressive approaches, but these approaches are limited to modeling the prior and/or likelihood models.

In this study, we introduce a neural ordinary differential equation (NODE~\cite{node}). formulation that describes 
an explicitly defined progressive SR procedure 
from the LR to HR images via a neural network.
In particular, we reconstruct the HR image by numerically solving the initial value problem originated from the proposed ODE formulation, given the LR image as an initial condition. 
With the aid of the proposed 
ODE, our method eases implementation using conventional restoration networks and ODE solvers without any exertion to improve the performance.
Furthermore, by simply changing the initial condition of our formulation at the test-time, ours can naturally handle a continuous-valued scale factor.
Extensive experiments demonstrate the superiority of the proposed method over state-of-the-art SR approaches.


\section{Proposed Method}


\begin{figure*}[ht]
\centering
\includegraphics[width=\linewidth]{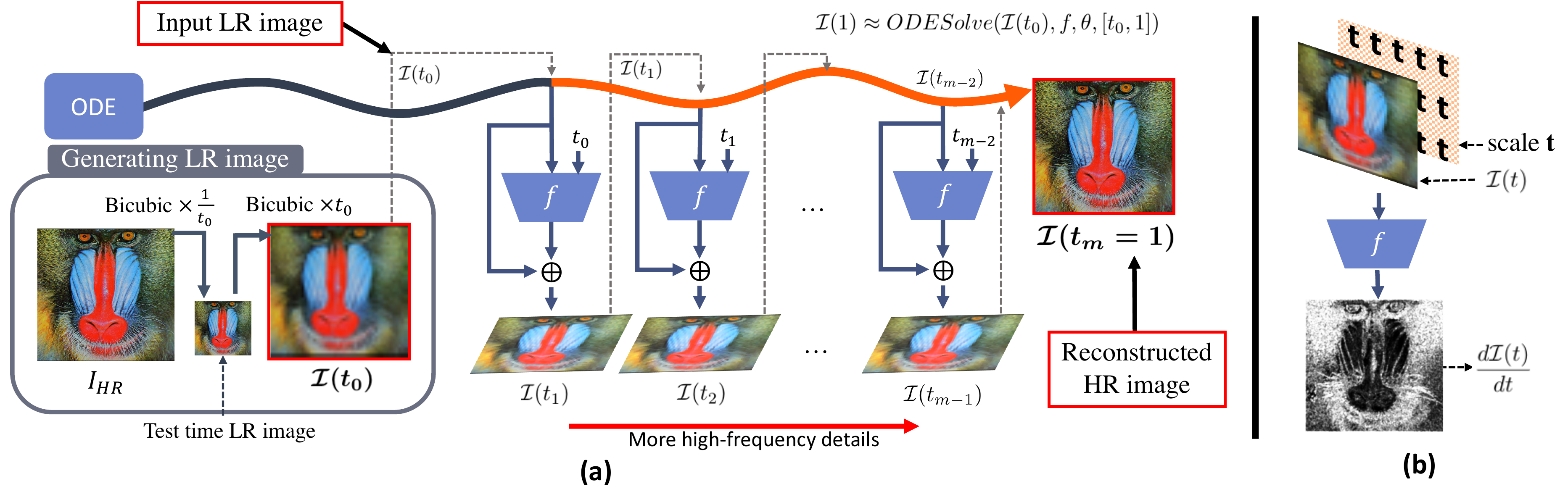}
\caption{(a) Overview of the proposed SR approach (NODE-SR). $\{t_i\}_{0 \leq i \leq m}$ is a strictly decreasing sequence and $t_m=1$. Solid orange line represents our SR process that starts with the initial condition $\mathcal{I}(t_0)$ until we reconstruct the final HR image $\mathcal{I}(1)$. (b) The neural network $f$ takes an input image $\mathcal{I}(t)$ with the scale factor $t$ and outputs the desired high-frequency detail. }
\label{explin_figure}
\end{figure*}

\subsection{Progressive Super-Resolution Formulation}

Existing SR methods utilizing progressive SR process~\cite{lapsr,dbpn,srfbn} are based on iterative multi-stage approaches and can be viewed as variants of the following:

\begin{equation}
I_{n} = g_{n-1} (I_{n-1}) \quad (n \leq N),
\label{eq_progressive_traditional}
\end{equation}
where $n$ denotes the iteration step, $I_0$ denotes the given initial input LR image, and $I_n$ is the iteratively refined image from its previous state $I_{n-1}$. These approaches typically produce multiple intermediate HR images during the refinement, and the rendered image at the last $N$-th iteration~\cite{drrn,srfbn} or a combined version of the multiple intermediate images ($\{ I_n\}_{1\leq n \leq N}$)~\cite{drcn,dbpn} becomes the final SR result. Although these previous progressive methods show promising SR results, they still have some limitations. 
First, these methods need plenty of time and effort in determining the network configurations including the number of progressive updates $N$ and hyper-parameter settings, and designing cost functions to train the SR networks $g$. In addition, well-engineered and dedicated learning strategy, such as curriculum learning~\cite{srfbn} and recursive supervision~\cite{drcn}, is required for each method. This complication comes from the lack of clear understanding on their intermediate image states $\{ I_n\}$. To alleviate these problems, we formulate the progressive SR process with a differential equation. This allows us to implement and train the SR networks in an established way while outperforming the performance of conventional progressive SR process.

Assume that $(I_{HR})\downarrow_t$ is a downscaled version of a ground-truth clean image $I_{HR}$ using a traditional SR kernel (e.g., bicubic) with a scaling factor $\frac{1}{t}$. We then define $\mathcal{I}(t)$ by upscaling $(I_{HR})\downarrow_t$ using that SR kernel with a scaling factor ${t}$ so that $I_{HR}$ and $\mathcal{I}(t)$ have the same spatial resolution (see the illustration of \say{Generating LR image} in \figref{explin_figure}(a)). Note that $t \geq 1$, and $\mathcal{I}(1)$ denotes the ground-truth clean image $I_{HR}$.
To model a progressive SR process,
we first estimate the high-frequency image residual with a neural network. Specifically, when $t$ is a conventional discrete-scaling factor (e.g., x2, x3, and x4), image residual between $\mathcal{I}(t)$ and $\mathcal{I}(t-1)$ can be modeled using a neural network $f_{\text{discrete}}$ as:

\begin{equation}
    \mathcal{I}(t-1) - \mathcal{I}(t) = 
    f_{\text{discrete}}(\mathcal{I}(t), t).
    \label{discrete}
\end{equation}
Notably, $\mathcal{I}(t-1)$ includes more high-frequency details than $\mathcal{I}(t)$ without loss of generality. In our method, we model the slightest image difference to formulate a continuously progressive SR process. Therefore, we take the scale factor $t$ to continuous domain, and reformulate \eqref{discrete} as an ODE with a neural network $f$ as:
\begin{equation}
    \frac{d\mathcal{I}(t)}{dt}= f(\mathcal{I}(t), t, \theta),
    \label{ode}
\end{equation}
where $\theta$ denotes the trainable parameter of the network $f$.
Using this formulation, we can predict the high-frequency image detail required to slightly enhance $\mathcal{I}(t)$ with the network $f$.
(Note that we can obtain $\mathcal{I}(t)$ with any rational number $t$ by adding padding to the border of image before resizing and then center cropping the image.)
As existing SR neural networks have been proven to be successful at predicting the high-frequency residual image~\cite{vdsr}, 
we can use conventional SR architectures as our network $f$ in \eqref{ode} without major changes.

\subsection{Single Image Super-Resolution with Neural Ordinary Differential Equation}
In this section, we explain how to super-resolve a given LR image with a continuous scaling factor using our ODE-based SR formulation in \eqref{ode}.

First, we obtain $\mathcal{I}(t_0)$ by upscaling the given LR input image (\say{Test time LR image} in \figref{explin_figure}(a)) using the bicubic SR kernel to a desired output resolution with a scaling factor $t_0$ . 
Next, we solve the ODE initial value problem in \eqref{ode} with the initial condition $\mathcal{I}(t_0)$ by integrating the neural network $f$ from $t_0$ to $1$ to acquire the high-quality image $\mathcal{I}(1)$ as follows:
\begin{equation}
\mathcal{I}(1) = \mathcal{I}(t_0) + \int_{t_0}^{1} f(\mathcal{I}(t), t, \theta) dt.
\label{odesolve}
\end{equation}

Specifically, we approximate the high-quality image $\mathcal{I}(1)$ given a fully trained neural network $f$, network parameter $\theta$, initial condition $\mathcal{I}(t_0)$, and integral interval $[t_0, 1]$ using an ODE solver ($ODESolve()$) as: 
\begin{equation}
\mathcal{I}(1) \approx ODESolve(\mathcal{I}(t_0), f, \theta, [t_0, 1]).
\label{eq_odelver}
\end{equation}
Thus, our method does not need to consider the stop condition (i.e., the number of feedback iterations) of the progressive SR process unlike conventional approaches~\cite{drrn,srfbn}.
Notably, during the training phase, we need to employ an ODE solver which allows end-to-end training using backpropagation with other components such as the neural network $f$.
Unlike other progressive SR methods~\cite{drcn,srfbn}, we do not require any other learning strategies like curriculum learning during the training phase.

\begin{figure}[ht]
\centering
\includegraphics[width=\linewidth]{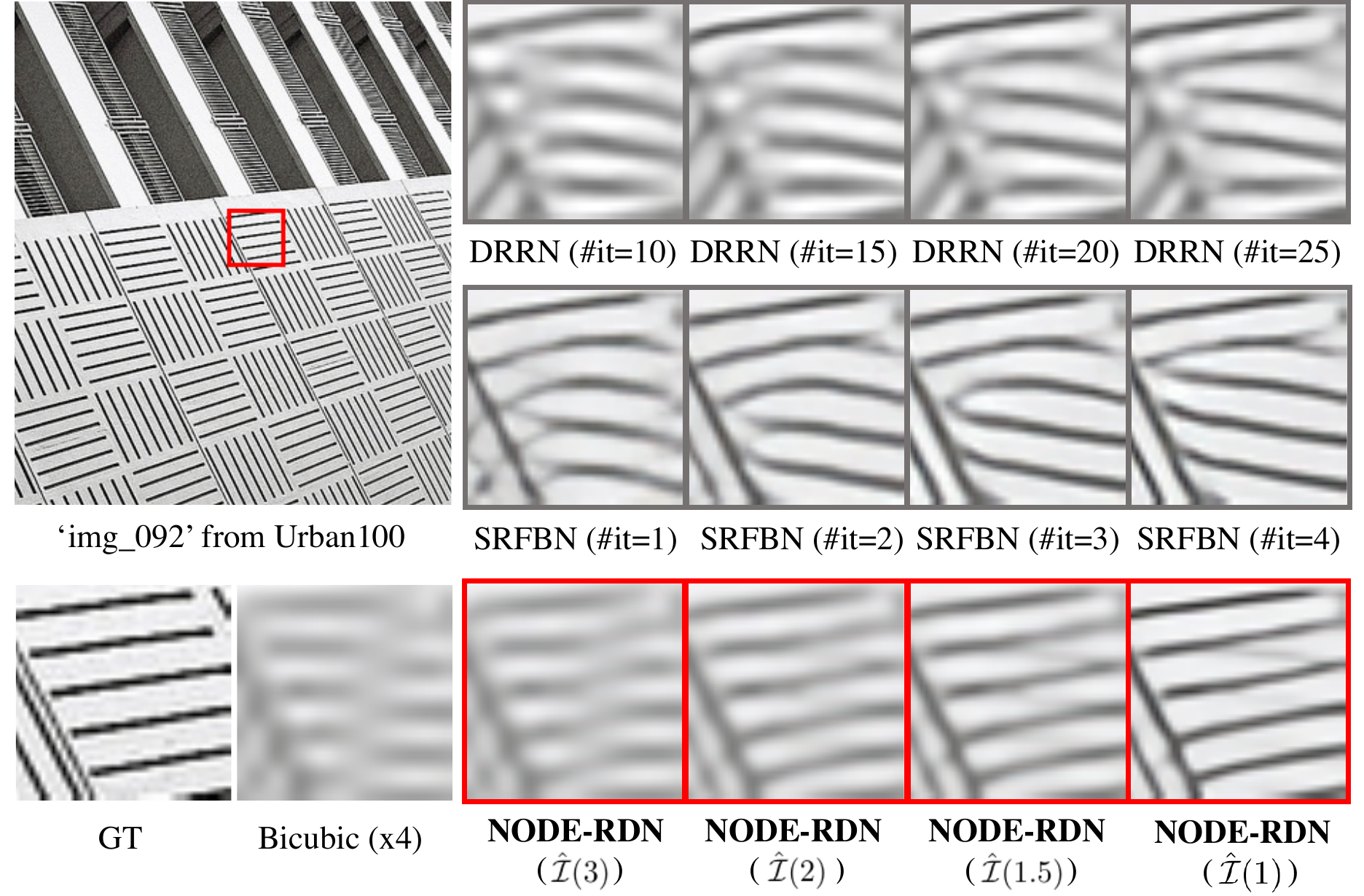}
\caption{Visual comparisons with conventional progressive SR methods (DRRN, SRFBN). For different scale factors (x2, and x4) intermediate HR images are visualized,
and \#it indicates the number of updates used to render results by DRRN and SRFBN. $\hat{I}()$ denotes the predicted results by our NODE-RDN.}
\label{progressive_change_qual}
\end{figure}

\begin{figure}[t]
    \centering
    \includegraphics[width=\linewidth]{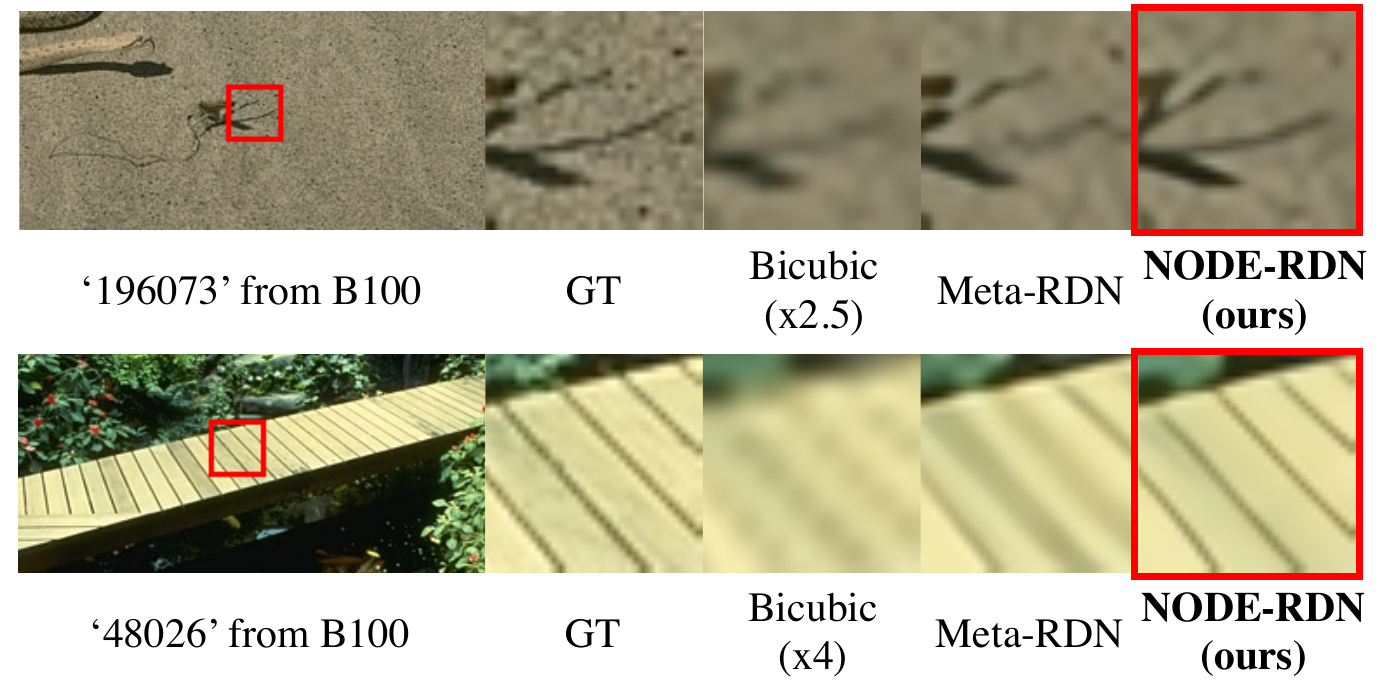}
    \caption{Visual comparison of NODE-RDN (ours) with Meta-RDN on scale x2.5 and x4.}
    \label{scale_arbitrary_qual}
\end{figure}

\noindent In addition, our formulation is made upon a continuous context, allows a continuous scale factor $t_0$ where $t_0 \geq 1$. 
This makes our method able to handle the arbitrary-scale SR problem.
To train the deep neural network $f$, and learn the parameter $\theta$ in \eqref{eq_odelver}, we minimize the loss summed over scale factors $t$ using the L1 loss function as:
\begin{equation}
    \mathcal{L}(\theta) = \sum_{t} \| I_{HR} - ODESolve(\mathcal{I}(t), f, \theta, [t, 1]) \|_1.
\label{loss}
\end{equation}

\noindent By minimizing the proposed loss function, our network parameter $\theta$ is trained to estimate the image detail to be added into the network input as in \eqref{ode}.

\section{Experimental Results}

In this section, we carry out extensive experiments to demonstrate the superiority of the proposed method, and add various quantitative and qualitative comparison results. We also provide detailed analysis of our experimental results.

\subsection{Implementation details}
\vspace{-5px}

We use VDSR~\cite{vdsr} and RDN~\cite{rdn} as backbone CNN architectures for our network $f$ with slight modifications.
For each CNN architecture, we change the first layer to feed the scale factor $t$ as an additional input. To be specific, we extend the input channel from 3 to 4, and the pixel locations of the newly concatenated channel (4-th channel) are filled with a scalar value $t$ as shown in \figref{explin_figure}(b). 
In addition, for RDN, we remove the last upsampling layer so that input and output resolutions are the same in our work. 
Note that, no extra parameters are added except for the first layers of the networks. 
To train and infer the proposed SR process, we use Runge–Kutta (RK4) method as our ODE solver in \eqref{loss}.
For simplicity, our approaches with VDSR and RDN backbones are called NODE-VDSR and NODE-RDN in the remaining parts of the experiments, respectively.
We use the DIV2K~\cite{div2k} dataset to train our NODE-VDSR and NODE-RDN. 
We train the network by minimizing the L1 loss in \eqref{loss} with the Adam optimizer ($\beta_1=0.9$, $\beta_2=0.999$, $\epsilon=10^{-8}$)~\cite{adam}. The initial learning rate is set as $10^{-4}$, which is then decreased by half every 100k gradient update steps, and trained for 600k iterations in total. The mini-batch size of NODE-VDSR is 16 (200x200 patches), but our NODE-RDN takes 8 patches as a mini-batch (130x130 patches) owing to the memory limit of our graphic units. Similar to the training settings in Meta-SR~\cite{metasr}, we train the network $f$ by randomly changing the scale factor $t$ in \eqref{loss} from 1 to 4 with a stride of 0.1 (i.e., $t \in \{1.1, 1.2, 1.3, ..., 4\}$).

\begin{table*}[ht]
\centering
\resizebox{\textwidth}{!}{
\begin{tabular}{|c|c|c|c|c|c|c|c|c|c|}
\hline
Dataset                   & Scale & Bicubic      & DRCN         & LapSRN      & DRRN         & PRLSR       & SRFBN        & NODE-RDN (ours) & NODE-RDN+ (ours)       \\ \hline
\multirow{3}{*}{Set14}    & x2    & 30.24/0.8688 & 33.04/0.9118 & 33.08/0.913 & 33.23/0.9136 & 33.69/0.9191 & 33.82/0.9196 & \textcolor{blue}{33.90}/\textcolor{blue}{0.9209}   & \textcolor{red}{33.95}/\textcolor{red}{0.9214} \\
                          & x3    & 27.55/0.7742 & 29.76/0.8311 & 29.87/0.833 & 29.96/0.8349 & 30.43/0.8436 & 30.51/0.8461 & \textcolor{blue}{30.53}/\textcolor{blue}{0.8465}   & \textcolor{red}{30.59}/\textcolor{red}{0.8473} \\
                          & x4    & 26.00/0.7027 & 28.02/0.7670 & 28.19/0.772 & 28.21/0.7721 & 28.71/0.7838 & \textcolor{blue}{28.81}/\textcolor{blue}{0.7868} & 28.76/0.7866   & \textcolor{red}{28.83}/\textcolor{red}{0.7877} \\ \hline
\multirow{3}{*}{B100}     & x2    & 29.56/0.8431 & 31.85/0.8942 & 31.80/0.895 & 32.05/0.8973 & 32.25/0.9005 & 32.29/0.9010 & \textcolor{blue}{32.34}/\textcolor{blue}{0.9025}   & \textcolor{red}{32.38}/\textcolor{red}{0.9028} \\
                          & x3    & 27.21/0.7385 & 28.80/0.7963 & 28.81/0.797 & 28.95/0.8004 & 29.14/0.8060 & 29.24/0.8084 & \textcolor{blue}{29.25}/\textcolor{blue}{0.8094}   & \textcolor{red}{29.28}/\textcolor{red}{0.8100} \\
                          & x4    & 25.96/0.6675 & 27.23/0.7233 & 27.32/0.728 & 27.38/0.7284 & 27.64/0.7378 & 27.72/0.7409 & \textcolor{blue}{27.72}/\textcolor{blue}{0.7410}   & \textcolor{red}{27.75}/\textcolor{red}{0.7417} \\ \hline
\multirow{3}{*}{Urban100} & x2    & 26.88/0.8403 & 30.75/0.9133 & 30.41/0.910 & 31.23/0.9188 & 32.35/0.9308 & 32.62/0.9328 & \textcolor{blue}{32.81}/\textcolor{blue}{0.9345}   & \textcolor{red}{32.97}/\textcolor{red}{0.9355} \\
                          & x3    & 24.46/0.7349 & 27.15/0.8276 & 27.06/0.827 & 27.53/0.8378 & 28.27/0.8541 & 28.73/0.8641 & \textcolor{blue}{28.81}/\textcolor{blue}{0.8644}   & \textcolor{red}{28.94}/\textcolor{red}{0.8662} \\
                          & x4    & 23.14/0.6577 & 25.14/0.7510 & 25.21/0.756 & 25.44/0.7638 & 26.22/0.7892 & \textcolor{blue}{26.60}/\textcolor{red}{0.8015} & 26.56/0.7985   & \textcolor{red}{26.68}/\textcolor{blue}{0.8010} \\ \hline
\end{tabular}
}
\caption{Comparison with progressive SR methods on the benchmark datsets (Set14~\cite{set14}, B100~\cite{b100}, and Urban100~\cite{urban100}). We provide average PSNR/SSIM values for scaling factors x2, x3, and x4. Our NODE-RDN and NODE-RDN+ show the best performance. Red and blue colors denote the best and second best results, respectively.}
\label{progressive_table}
\end{table*}

\begin{table*}[ht]
\centering
\resizebox{\textwidth}{!}{
\begin{tabular}{|l|c|c|c|c|c|c|c|c|c|c|c|c|c|c|c|}
\hline
\multicolumn{1}{|c|}{\diagbox[width=10em]{Methods}{Scale}} & x1.1           & x1.2           & x1.3           & x1.4           & x1.5           & x1.6           & x1.7           & x1.8           & x1.9           & x2.0           & x2.1           & x2.2           & x2.3           & x2.4           & x2.5           \\ \hline
bicubic                             & 36.56          & 35.01          & 33.84          & 32.93          & 32.14          & 31.49          & 30.90          & 30.38          & 29.97          & 29.55          & 29.18          & 28.87          & 28.57          & 28.31          & 28.13          \\ \hline
VDSR                                & -              & -              & -              & -              & -              & -              & -              & -              & -              & 31.90          & -              & -              & -              & -              & -              \\
VDSR+t                              & 39.51          & 38.44          & 37.15          & 36.04          & 34.98          & 34.15          & 33.39          & 32.78          & 32.22          & 31.70          & 31.27          & 30.86          & 30.53          & 30.2           & 29.91          \\
NODE-VDSR (ours)                     & \textbf{41.46} & \textbf{39.36} & \textbf{37.75} & \textbf{36.51} & \textbf{35.38} & \textbf{34.49} & \textbf{33.70} & \textbf{33.07} & \textbf{32.50} & \textbf{31.95} & \textbf{31.52} & \textbf{31.09} & \textbf{30.76} & \textbf{30.42} & \textbf{30.12} \\ \hline
RDN                                 & -              & -              & -              & -              & -              & -              & -              & -              & -              & 32.34          & -              & -              & -              & -              & -              \\
RDN+t                               & 42.83          & 39.92          & 38.18          & 36.87          & 35.71          & 34.80          & 33.99          & 33.34          & 32.77          & 32.22          & 31.76          & 31.33          & 30.99          & 30.64          & 30.34          \\
Meta-RDN                            & 42.82          & 40.04          & 38.28          & 36.95          & 35.86          & 34.90          & 34.13          & 33.45          & 32.86          & 32.35          & 31.82          & 31.41          & 31.06          & 30.62          & 30.45          \\
NODE-RDN (ours)                      & 43.22          & 40.06          & 38.35          & 37.02          & 35.86          & 34.95          & 34.14          & 33.47          & 32.89          & 32.34          & 31.89          & 31.46          & 31.12          & 30.76          & 30.46          \\
NODE-RDN+ (ours)                     & \textbf{43.33} & \textbf{40.13} & \textbf{38.40} & \textbf{37.07} & \textbf{35.90} & \textbf{34.99} & \textbf{34.17} & \textbf{33.50} & \textbf{32.93} & \textbf{32.38} & \textbf{31.93} & \textbf{31.50} & \textbf{31.16} & \textbf{30.80} & \textbf{30.50} \\ \hline

\end{tabular}
}
\\
\resizebox{\textwidth}{!}{

\begin{tabular}{|l|c|c|c|c|c|c|c|c|c|c|c|c|c|c|c|}
\hline

\multicolumn{1}{|c|}{\diagbox[width=10em]{Methods}{Scale}} & x2.6           & x2.7           & x2.8           & x2.9           & x3.0           & x3.1           & x3.2           & x3.3           & x3.4           & x3.5           & x3.6           & x3.7           & x3.8           & x3.9           & x4.0           \\ \hline
bicubic                             & 27.89          & 27.66          & 27.51          & 27.31          & 27.19          & 26.98          & 26.89          & 26.59          & 26.60          & 26.42          & 26.35          & 26.15          & 26.07          & 26.01          & 25.96          \\ \hline
VDSR                                & -              & -              & -              & -              & 28.83          & -              & -              & -              & -              & -              & -              & -              & -              & -              & 27.29          \\
VDSR+t                              & 29.64          & 29.39          & 29.15          & 28.93          & 28.74          & 28.55          & 28.38          & 28.22          & 28.05          & 27.89          & 27.76          & 27.58          & 27.47          & 27.34          & 27.20          \\
NODE-VDSR (ours)                     & \textbf{29.85} & \textbf{29.61} & \textbf{29.36} & \textbf{29.14} & \textbf{28.94} & \textbf{28.75} & \textbf{28.58} & \textbf{28.41} & \textbf{28.25} & \textbf{28.08} & \textbf{27.96} & \textbf{27.79} & \textbf{27.66} & \textbf{27.54} & \textbf{27.40} \\ \hline
RDN                                 & -              & -              & -              & -              & 29.26          & -              & -              & -              & -              & -              & -              & -              & -              & -              & 27.72          \\
RDN+t                               & 30.06          & 29.80          & 29.55          & 29.33          & 29.12          & 28.92          & 28.76          & 28.59          & 28.43          & 28.26          & 28.13          & 27.95          & 27.84          & 27.71          & 27.58          \\
Meta-RDN                            & 30.13          & 29.82          & 29.67          & 29.40          & \textbf{29.30} & 28.87          & 28.79          & 28.68          & 28.54          & 28.32          & 28.27          & 28.04          & 27.92          & 27.82          & 27.75          \\
NODE-RDN (ours)                      & 30.18          & 29.93          & 29.67          & 29.45          & 29.25          & 29.05          & 28.88          & 28.71          & 28.54          & 28.37          & 28.24          & 28.07          & 27.96          & 27.81          & 27.72          \\
NODE-RDN+ (ours)                     & \textbf{30.22} & \textbf{29.97} & \textbf{29.71} & \textbf{29.49} & 29.28          & \textbf{29.05} & \textbf{28.92} & \textbf{28.74} & \textbf{28.58} & \textbf{28.41} & \textbf{28.28} & \textbf{28.12} & \textbf{28.00} & \textbf{27.87} & \textbf{27.75} \\ \hline
\end{tabular}
}

\caption{Average PSNR values on the B100~\cite{b100} evaluated with different scale factors. The best performance is shown in \textbf{bold number}. }
\label{scale_arbitrary_table}
\end{table*}

\vspace{-15px}
\subsection{Comparison with Progressive SR Methods}
\vspace{-5px}

First, we compare our NODE-RDN with several state-of-the-art progressive SR methods: DRCN~\cite{drcn}, LapSRN~\cite{lapsr}, DRRN~\cite{drrn}, PRLSR~\cite{prlsr}, and SRFBN~\cite{srfbn}. As in~\cite{edsr}, self-ensemble method is used to further improve NODE-RDN (denoted as NODE-RDN+). 
Note that, our NODE-RDN and NODE-RDN+ can handle multiple scale factors $t$ including non-integer scale factors (e.g., x1.5) using the same network parameter. In contrast, other approaches are required to be trained for certain discrete integer scale factors (x2, x3, and x4) separately, resulting in a distinct parameter set for each scale factor.
Nevertheless, quantitative restoration results in \tabref{progressive_table} show that our NODE-RDN, NODE-RDN+ consistently outperforms conventional progressive SR methods for the discrete integer scaling factors (x2, x3, and x4) in terms of PSNR.
In \figref{progressive_change_qual}, we investigate intermediate images produced during the progressive SR process with the scale factors x2 and x4. Final results by DRRN are obtained after 25 iterations, and the final results by SRFBN are obtained with 4 iterations as in their original settings. We provide 4 intermediate HR images during the updates for visual comparisons.
For our NODE-RDN, intermediate image states are represented as $\hat{\mathcal{I}}(t_i)$ where ${1 \leq t_i \leq t_0}$ and
$\hat{\mathcal{I}}(t_i) = ODESolve(\mathcal{I}(t_0), f, \theta, [t_0, t_i])$.
We observe that DRRN and SRFBN fail to progressively refine patches with high-frequency details, while our NODE-RDN can gradually improve the intermediate images and render promising results at the final states.

\vspace{-15px}
\subsection{Comparison with Multi-scale SR Methods}
\vspace{-5px}

Our approach can handle a continuous scale factor for the SR task, thus we compare ours with existing multi-scale SR methods that can handle continuous scale factors:
VDSR~\cite{vdsr} and Meta-SR~\cite{metasr}.
Notably, Meta-SR implemented using RDN (i.e., Meta-RDN) is the current state-of-the-art SR approach. 
In \tabref{scale_arbitrary_table}, we show quantitative results compared to existing SR methods (VDSR, RDN, and Meta-RDN).
Note that, VDSR+t and RDN+t are modified versions of VDSR and RDN to take the scale factor $t$ as an additional input of the networks and have the same input and output resolutions as in our network $f$.
We also compare our method with these new baselines (VDSR+t and RDN+t) for fair comparisons.
We evaluate the SR performance on the B100 benchmark dataset by increasing the scaling factor from 1.1 to 4. 
Interestingly, we observe that NODE-VDSR outperforms VDSR and VDSR+t at every scale by a large margin although VDSR and VDSR+t have similar network architecture to our NODE-VDSR. 
Similarly, NODE-RDN shows better performance than Meta-RDN and RDN+t. 
We also provide qualitative comparison results with Meta-SR in \figref{scale_arbitrary_qual}, and we see that 
our NODE-RDN recovers much clearer edges than Meta-RDN.

\vspace{-10px}
\section{Conclusion}
\vspace{-3px}
In this work, we proposed a novel differential equation for the SR task to progressively enhance a given input LR image, and allow continuous-valued scale factor. Image difference between images over different scale factors is physically modeled with a neural network, and formulated as a NODE. To restore a high-quality image, we solve the ODE initial value problem with the initial condition given as an input LR image. The main difference with existing progressive SR methods is that our formulation is based on the physical modeling of the intermediate images, and adds fine high-frequency details gradually. 
Detailed experimental results show that our method achieves superior performance compared to state-of-the-art SR approaches.

\bibliographystyle{IEEEbib}
\bibliography{strings,refs}

\end{document}